\documentstyle[pra,aps,floats,epsfig,twocolumn]{revtex}

\def\ang{\mbox{\AA}}
\def\mev{\mbox{meV}}
\def\nm{\mbox{nm}}
\def\be{\begin{equation}}
\def\ee{\end{equation}}
\def\bea{\begin{eqnarray}}
\def\eea{\end{eqnarray}}

\begin{document}
\title{Transition from  reflection to sticking in ultracold atom-surface
scattering}

\author{Areez Mody, John Doyle and Eric J. Heller}

\date{August 2000}

\address{Department of Physics, Harvard University, Cambridge, MA 02138}

\maketitle

\begin{abstract}
In paper I \cite{paper1} we showed that under very general circumstances, atoms
approaching a
surface will not stick as its incoming energy approaches zero. This is true
of either warm
or cold surfaces. Here we explore the transition region from non sticking to
sticking as energy is increased. The key to
understanding the transition region is the WKB approximation and the nature
of its
breakdown. Simple rules for understanding the rollover to higher energy,
post-threshold
behavior, including analytical formulae for some asymptotic forms of the
attractive
potential are presented.  We discuss a practical example of atom-surface pair
in various substrate geometries.  We also discuss the case of low
energy scattering from clusters.
\end{abstract}

\section{Introduction}
\label{} 

The problem of sticking of atoms to surfaces at very low collision
velocities has a long history and has met with some controversy. The issue
goes back
to the early distorted wave Born approximation results of
Lennard-Jones \cite{lj}, who
obtained the threshold law sticking probability going as $k$ in the
limit of low
velocities. 
This paper is a companion to paper I \cite{paper1},
 wherein we put the problem on a
firmer theoretical foundation. We showed (non-perturbatively) that in
an ultracold collision a simplistic one-body view of things is
essentially correct even if the number of internal degrees of freedom is
very large.
  We concluded that approaching atoms will not stick to surfaces if the
approach velocity is low enough, even if the surface is warm. From the
methods used, it is clear that the non-sticking rule would apply to
clusters as well as semi-infinite surfaces, and would also apply to
projectiles more
complex than atoms.

From an experimental perspective atom-surface sticking could impact the area of
 guiding
and trapping atoms in material wires and containers. In those
 applications
it is necessary
to predict
the velocities needed for quantum
reflection, sticking, and the transition regime
between them. We do so in this paper.

Above a certain temperature or kinetic energy, but still well below the
attractive well
depth of the atom surface potential, atoms will stick to surfaces with near
100 percent
certainty. The reason for this is simple: Classical trajectory
simulations of atom-surface collisions at low collision velocities
indicate sticking with 
near certainty
because the acceleration in the attractive regime is followed by a hard
collision with the
wall. This almost always leads to sufficient energy loss from 
the particle to the
surface that
immediate escape is not possible. This is true so long as the approach
energy is
significantly less than the well depth, which is itself greater than the
temperature of the surface.
Therefore, the onset of quantum reflection is heralded by a break down in
the WKB
approximation - an approximation based purely on the (sticking)
classical trajectories.

Thus, there must exist a transition region between the non-sticking regime
for
very low collision velocities, and the sticking regime at higher velocities.
The key to
understanding the transition region is to understand the validity of
classical mechanics (WKB)
as applied to the sticking problem.
 The correctness of the simplistic one-body
physics of quantum reflection from the surface, focusses our
study  on the WKB approximation to  the coordinate normal to the
surface.  The entrance channel wavefunction thus obtained may also be used
as input
into the Golden Rule to study the threshold behaviour of the inelastic
cross-sections.
We do this in Section 
\ref{relation_to_threshold_behaviour}. The nonsticking threshold
behavior we established in paper I  is interpreted as an extension of
the validity of the so-called Wigner threshold behavior. We are also able to
make definite
predictions about the nature of the post threshold behavior of sticking in
terms of
inelastic cross-sections. (Section
\ref{relation_to_threshold_behaviour}).

\section{Quantum reflection and WKB}
\label{quantum_reflection_and_WKB}
We consider the typical case of an attractive potential arising out
of the cumulative effect of Van der Waals attractions between target
and incident atoms.  A classical atom would proceed
straight into the interaction region showing no sign of reflection,
but the quantum mechanical probability of being found inside is suppressed
by a factor of $k$ (as $k \rightarrow 0$) as compared to the
classical probability (Section \ref{incidence_on_slab} ),
where
 $k$ is
the wave
vector of the incoming atom. This is
tantamount to saying that quantum
mechanically the amplitude is reflected back without penetrating the
interaction region, analogous to the elementary case of reflection
from the edge of a step-down potential in one dimension while
attempting
to go over the edge.
 A useful way to view this is to attribute the reflection to
the failure of the WKB approximation.

To be specific, we keep the geometry of paper I in mind: an atom is
incident from the right ($x>0$) upon the face of a slab ($x=0$) that
lies to the left of $x=0$.
For a low incoming energy $\epsilon \equiv \hbar^2 k^2/2m$, a left-moving
WKB
solution begun well inside the interaction region will fail to match
onto a purely left-going WKB solution as we integrate out to large
distances
because the WKB criterion
\begin{equation}
|\lambda'(x)| = \left|{\hbar p' \over p^2}\right| \ll 1
\label{wkb_error}
\end{equation}
for the local accuracy of the wavefunction will in general fail to
be valid in some intermediate region. For bounded potentials that turn on
abruptly for example at $x=a$, it is obvious that WKB will fail 
near $x \sim a $.
For long-range potentials such as a power law $V(x) = -c_n/x^n$
it is not immediately obvious where this region of WKB failure lies,
if it exists at all.
It turns out that even in this case it is possible to identify (for small
enough $\epsilon$) a distance (dependent on $\epsilon$) at which the
potential `turns on' and where WKB will fail.  We will show below
that WKB is at its worst ($|\lambda'(x)|$ is maximized) at a
distance $x$ where the kinetic and potential energies are
approximately equal, i.e  where $|V(x)| \sim \epsilon$.
The distance away from the slab at which the particle is turned
around - or quantum reflected - is precisely this distance.  Furthermore,
one may heuristically expect that the greater the failure
of WKB, the greater the reflection.

\begin{figure}
\centerline{\epsfig{figure=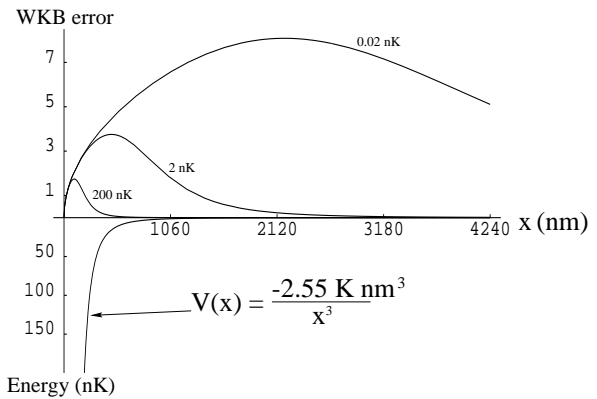,width=0.9\hsize}}
\hspace{0.1in}
\caption{
The WKB error of Eq.~(\ref{wkb_error}) 
for three different
values of the incoming energy 200, 2 and 0.02 nK,
vs. the distance $x$ nm from the slab (SiN).  The long
range form of the potential $-c_3/x^3$ ($c_3$=220 $\mev \; \ang^3$) 
is also shown 
for which the negative `y-axis' is calibrated in the different units
of energy.  The sticking probabilities for the three cases are
approximately 1, 0.6, 0.1.}
\label{fig:wkb_error}
\end{figure}

Fig. \ref{fig:wkb_error}
shows a plot of the error term in  
Eq.~(\ref{wkb_error})
 for three different 
values of the incoming energy of neon on a semi-infinite slab of SiN.
 The essential points to
notice are: \newline
1) There is a greater error incurred in attempting to apply the WKB
(classical mechanics) approximation to colder atoms than to warmer
ones.  Consequently, we expect that the slower the atom, the more
non-classical its behavior.  In particular, slow enough atoms will be
`quantum reflected' and will not stick. \newline
2) As the incoming velocity is decreased the atom is reflected at
distances progressively further and further away from the slab.  This
is because the interval in $x$ around which the WKB error is large, may
be identified as the region from which the atom is reflected.  

A useful qualitative rule of thumb 
obtained in Section \ref{sec:WKB_fail} below
is that the region of WKB error
reaches all the way out to those regions where the potential energy is
still roughly the same order of magnitude as the incoming energy 
(Eq.~(\ref{V_equal_8e})).
This means that as $\epsilon \rightarrow 0$ the error is still large 
where the potential energy graph looks essentially flat.  In fact 
as $\epsilon \rightarrow 0$ it is easily
shown that a plot of the WKB error will show a
non-uniform convergence to a polynomial proportional to 
\begin{equation}
x^{{n \over 2}-1} \hspace{1cm} \mbox{for  all} \;  n>0
\label{polynomial}
\end{equation}
Fig. \ref{fig:wkb_error} shows the case for $n=3$.

\section{WKB failure}
\label{sec:WKB_fail}
Differentiating $p^2 / 2m + V(x) = \epsilon$  w.r.t. x, we have
\begin{equation}
p' = {-m V' \over p}
\end{equation}
which when used repeatedly to eliminate $p'$ shows that $|p' / p^2|$
in Eq.~(\ref{wkb_error})
is maximized when 
\begin{equation}
{p^2 \over 3m} = {V'^2 \over V''}.
\end{equation}
For $V(x) = - c_n/x^n$, this is
exactly when 
\begin{equation}
|V(x)| = \epsilon \left({2(n+1) \over n-2} \right).
\label{V_equal_8e}
\end{equation}
We discover that for $n>2$ only, we have a point
where WKB is at its worst at a distance $x$ where
$|V(x)| \sim \epsilon$, and moreover, that this maximum behaves like
\begin{equation}
\label{max_error}
 \mbox{max}  \left| {p^{\prime} \over p^2} \right| \sim
 {1 \over c_n^{1/n} \,\epsilon^{ {1 \over 2} - {1 \over n}}} \sim
 {1 \over c_n^{1/n} \,k^{1-{2 \over n}}}
\end{equation}
which for $n>2$ diverges as $k \rightarrow 0$.
Note how a {\it weaker} potential (smaller $c_n$) is {\it better} at
reflecting a particle
at the same energy, but allows the atom to approach closer.
Heuristically a sketch of $V(x) = - c_n/x^n$ reveals
why:  the weaker potential is seen to turn on more abruptly at a point
closer to $x=0$, promoting
an greater breakdown of WKB there.  Alternatively a simple scaling argument
with Schrodinger's equation reveals the same trend.

The above conclusions are valid only for $n > 2$.  For $n \leq 2$ the
error term of Eq.~(\ref{wkb_error}) looks qualitatively different from that in
Fig. \ref{fig:wkb_error}.
It is small at all distances except near $x = 0$ where it diverges to
infinity, as is evident from Eq.(\ref{polynomial}).  If the physical
parameters are such that this region where WKB fails very close
to the slab is never actually manifest in the long-range part of the 
potential then the `no-reflection' classical behaviour will be valid
all the way up to distances near the slab where the atom will begin
to feel the short range forces and loose energy to the internal
degrees of freedom.
For such a case then with $n<2$ we believe one will {\em not} 
observe quantum reflection.
 
\section{Sticking probability}
Having established that the reflection is caused by a well-defined
localized region, we solve the one-dimensional
Schrodinger equation around this region to accurately compute the
reflection probability.
For an attractive power law potential $V(x) = - c_n/x^n$, 
the relevant one dimensional
equation is
\begin{equation}
 \left({d^2 \over dx^2} + {a_n^{n-2} \over x^n}
   + k^2 \right) \phi_e(x) = 0.
\label{1d_equation}
\end{equation}
$\phi_e(x)$ is the entrance channel wavefunction.
The length scale
\begin{equation}
 a_n\equiv {(2 m c_n / \hbar^2)}^{1 \over n-2},
\label{a_n}
\end{equation}
contains all the qualitative information about the reflection. Its
relevance is twofold. Firstly, the sticking probability for small $k$,
behaves as 
\begin{equation}
P_{\rm sticking} \sim N_n \, k \ a_n
\label{P_sticking}
\end{equation}
where $N_n$ is a pure numeric constant (roughly of order 10 for $n=
3$, and of order 1 for $n = 4,5$), see Ref. \cite{cote}.
$P_{\rm sticking}$ may be computed numerically for any $k$, and Fig.2 shows
$P_{\rm sticking}$ vs. $k a_n$ for $n=3,4$, and 5.
Secondly, the distance at which the particle is turned
around is estimated by solving
\begin{equation}
\left({a_n \over x}\right)^n = (k a_n)^2
\label{V_equal_e}
\end{equation}
for x, which is just the requirement that $|V(x)| = \epsilon$.
Equation (\ref{P_sticking}) together with equation (\ref{a_n})
makes plain that a smaller $c_n$ is more
conducive to making quantum reflection happen, while Eq.~(\ref{V_equal_e})
indicates that the turnaround point is then necessarily closer to the
surface.  With these effects in mind, we look at some specific cases.

\begin{figure}
\centerline{\epsfig{figure=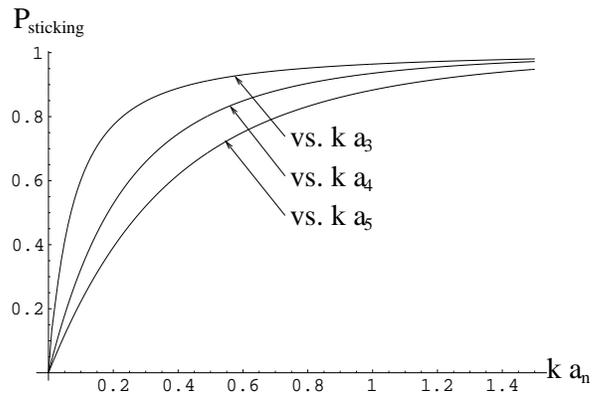,width=0.9\hsize}}
\hspace{0.1in}
\caption{Sticking probabilities for an atom incident on surface 
providing a long range interaction of the form $V(x)=-c_n/x^n$ for
the cases n=3,4,5. Note that the length scale $a_n$ used to compute
the dimensionless $k a_n$ coordinate on the `x-axis' vs. which
we plot the sticking probabilities, is different for
each $n$ . 
}
\label{fig:general_sticking_probabilities}
\end{figure}

\section{Examples}
We examine the case of incidence on a slab which may be treated as
semi-infinite, and also the case when it is a thin film.  It is useful to
first look at these cases  pretending there is
no Casimir interaction, and assuming that the short range
form of the potential is everywhere valid.  Afterwards we put in the
Casimir interaction.  For clarity we will pick a specific example of target and
incident
atoms for most of our discussions,
 by specifying the numeric values for the short range
potential between the atom and semi-infinite slab, since
these are most comprehensively tabulated in reference
\cite{c3_coefficients}.

\begin{figure}
\centerline{\epsfig{figure=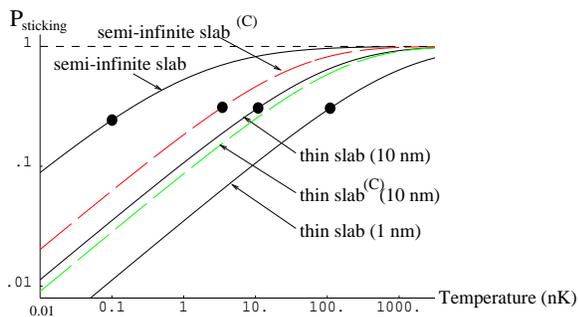,width=0.9\hsize}}
\hspace{0.1in}
\caption{Sticking probabilities vs temperature of incident Ne atoms
on SiN.
The broken line indicate the inclusion of the very long range Casimir 
forces (see text). The large dot demarcates the regions of threshold
and post-threshold, using the criterion suggested at the end of 
Section \ref{sec:WKB_failure}
}
\label{fig:example_sticking_probabilities_loglog}
\end{figure}

Fig. \ref{fig:example_sticking_probabilities_loglog}
shows the sticking
 probability
 vs. the temperature of an
incoming Ne atom in units of $10^{-9}$ Kelvin.   The slab is 
silicon-nitride (SiN). The various curves are for the different
cases depending on whether we are we are considering a thick
or thin slab, and whether the
Casimir effect is included or not. We will discuss these cases below, 
pointing out the relevant length and energy scales involved in
deciding
to label the slab as semi-infinite or thin.
The mapping from the mathematically natural $k a_n$ (with $n=3$ 
and $c_3=220 \ \mev \; \ang^3$) scale of 
Fig. \ref{fig:general_sticking_probabilities} to
the more physical temperature scale of
 Fig. \ref{fig:example_sticking_probabilities_loglog} is made using 
\begin{equation}
T \simeq \ 
[69.08  \ \ \mbox{Kelvin}] \ \ 
{\left( {m_H \over m_{\rm atom}} \right )}^3 
 \left({\mev \;\ang^3\over c_3}\right)^{2} \, (k a_3)^2
\label{T_equation}
\end{equation}
where we used
$\langle \epsilon \rangle = (3/2)k_B T$ to compute the temperature by
setting\label{sec:WKB_failure}
$\langle \epsilon \rangle$ equal to the incoming energy. 
$m_H =$ mass of hydrogen atom, and for our example $m_{\rm Ne}=20.03 \; m_H$.

All the graphs in 
 Fig. \ref{fig:example_sticking_probabilities_loglog}
have an initial slope of 0.5 indicating the $\sqrt{\epsilon}$
behaviour
of the sticking probabilities once the energies are low enough to be
in the Quantum Reflection regime.  A particular temperature at
which there is a transition to the post-threshold sticking regime,
we arbitrarily (but intuitively) define as the temperature where the slope becomes
0.4.  For the thin film case of 10 $nm$ in our example this
temperature is 10 nK.

While the parameters in our example are fairly typical, it is clear
 that
the cubic dependence on mass and quadratic dependence on the $c_3$
 coefficient in Eq.(\ref{T_equation}), will make this temperature
 range over quite a few orders of magnitude.  The $c_3$'s in Ref
\cite{c3_coefficients}, listed in units of $\mev-\ang^3$ for a variety of
surface
atom pairs, range in values from 100 to 3000.

\subsection{Semi-Infinite Slab (without Casimir)}

Even though $c_3$ coefficients are known both theoretically and 
experimentally for many surface-atom pairs, for completeness
 we take a moment to look at
a quick way of estimating them.
  This is provided by 
the London formula
\begin{equation}
 V_{\rm atom-atom}(r) = {-3 \over 2}{I_A I_B \over I_A + I_B}
   {\alpha_A \alpha_B \over r^6} \equiv {-c_6 \over r^6},
\label{London_formula}
\end{equation}
which estimates the Van der Waals interaction between any two
atoms.  $I$ is the ionization potential and $\alpha$ the polarizability
of each atom.  Then summing over all the atoms in the semi-infinite
slab (thick) we get
\begin{equation}
 V_{\rm slab-atom}(x) = {-\pi c_6 \rho_{\rm atoms} \over 6}
      \times {1 \over x^3}
 \equiv {-c_3 \over x^3 }
\label{slab_atom_interaction}
\end{equation}
where $\rho_{\rm atoms} =$ the density of slab atoms.
These estimates are not very accurate, but correctly indicate
the physical quantities on which the answer depends.  Reference
\cite{c3_coefficients} provides
a useful compendium of these coefficients.  We have used
$c_3=220 \pm 4 \ 
\mev \; \ang^3$ for neon atoms incident on silicon nitride from
work of 
\cite{c3_value}.  This
choice of $c_3$
makes
\begin{equation}
a_3 \simeq 212 \ \nm
\end{equation}
Thus the `semi-infinite slab' curve of Fig. 
\ref{fig:example_sticking_probabilities_loglog} is the $n=3$ curve of
Fig. \ref{fig:general_sticking_probabilities} scaled to temperature units
using
Eq. (\ref{T_equation}).

\subsection{Thin Slab (without Casimir)}
From far enough away any slab will appear thin.  
The surface-atom
interaction will behave like
\begin{equation}
{-c_3 \over x^3} - {-c_3 \over (x+d)^3} \simeq
{-3dc_3 \over x^4}
\label{thin_slab_pot}
\end{equation}
for $x \gg d$, where d is the thickness of the slab.
The resulting $c_4$ coefficient equal to $3 d c_3$ gives an $a_4 $
coefficient
that can be written as
\begin{equation}
a_4=\sqrt{{2m \over \hbar^2} 3 d c_3}=a_3 {\left[{3 d\over
      a_3}\right]}
^{(1/2)}.
\label{a_4_eqn}
\end{equation}
For macroscopic values of $d (\gg a_3)$ then, it is only for vanishingly
small incident energies that the finiteness of the slab becomes
apparent.  For any macroscopic $d$ this will be physically irrelevant.
For microscopic $d (\ll a_ 3) $ however, this window in energy over which the
thinness of the slab makes an appreciable difference
can be larger and even prevail for all energies.  To continue our
illustrative example we pick the microscopic value of $d = 10$ nm.
This makes $a_ 4 \simeq 800 \nm$.
The `thin slab' curve of Figure
\ref{fig:example_sticking_probabilities_loglog}
 shows
that the sticking probabilities are substantially reduced and the
onset of Quantum Reflection occurs at a much higher energy.

As a benchmark case, we also include what will likely be the physically
limiting case for a continuous film of $d=1$ nm. This further reduces the sticking
probabilities for a fixed temperature by a factor of 
 $\sqrt{10}$, because the important quantity $a_4$ is reduced by this
 much.
(Eq.(\ref{a_4_eqn})).  The transition temperature appears to have 
increased by 3 orders of magnitude versus the semi-infinite case.

\subsection{Semi-infinite slab (Casimir Regime)}
As the incoming energy $\epsilon$ tends to 0, we have seen that the
turn-around region from which the atom `quantum reflects' moves
progressively further away from the slab.  But at large distances,
however, it is well known that the interaction potential itself takes
on a different form due to Casimir effects.  In particular, a
semi-infinite dielectric slab (dielectric constant $\epsilon_s$) has an
interaction potential with an
atom of polarizability $\alpha$ given by \
\begin{eqnarray}
V_{\rm slab-atom}(x)&=& {-3 \over 8\pi} {\hbar c \alpha \over x^4}
{\epsilon_s - 1 \over \epsilon_s + 37/23}\hspace{1.7cm}x \rightarrow
\infty \\
&=& {-235 (eV-\ang) \alpha \over x^4} 
{\epsilon_s - 1 \over \epsilon_s + 37/23}
\hspace{.3cm}x \rightarrow \infty
\label{casimir_slab_atom}
\end{eqnarray}
Even for sufficiently large $x$, the form above is not exact but a
good approximation found in Ref. \cite{Casimir_estimates}.  Our
purpose here is only to estimate the
various numbers to see their relevance.  It will suffice to put 
$\alpha_{Ne}=0.39 \ang^3$ and
the last factor involving $\epsilon_s$ is replaced by 1 since most
solids and liquids have $\epsilon_s$ substantially greater than 1.  This
gives a $c_4^{(C)} $ coefficient of $9 \times 10^4 \mev \ang^4$ and hence a
resulting $a_4^{(C)}=93$ nm. 
The superscript `C' reminds us it is due to the Casimir
interaction  which is valid only for large enough $x$.

To estimate the distance beyond which the Casimir form itself is
valid, we use the statement from Ref.\cite{Hinds}: `Within a factor of 2,
the
van der Waals potential is correct at distances less than
0.12$\lambda_{tr}$, while the Casimir potential is correct at distances at
longer range.'  $\lambda_{tr} = [1,240 \, \nm] ({eV \over \Delta E})$ here is the
wavelength associated with the
transition between the ground and excited state that gives the atom
its polarizability.  $\Delta E$ is the transition energy. Knowing this
much we may deduce the qualitative features of the sticking
probability  
curve the arguments being similar to the cases above.

For this Casimir case and the one below, however, there is a caveat to 
all this.  The exact
manner in which the potential changes its  near range form to its
long-range Casimir form can certainly affect the sticking
probabilities at the intermediate energy where it makes this transition.
Some numerical experimentation choosing arbitrary forms of the
potential having the correct short range and long-range behavior,
confirms this.  Therefore the curves in figure 3 involving Casimir
forces are only quantitatively and {\em not} quantitatively correct.

\subsection{Thin Slab (Casimir Regime)}
Even for a thin slab we expect that the distance at which the Casimir
interaction is valid
remains the same as for a semi-infinite slab made of the same
material.  At these distances  if
$x \gg d$ is also valid,
then one may expect the surface atom interaction to behave like
\begin{equation}
{-c_4^{(C)} \over x^4} - {-c_4^{(C)} \over (x+d)^4} \simeq
{-4dc_4^{(C)} \over x^5}
\end{equation}
The length scale 
\begin{equation}
a_5^{(C)} = a_4^{(C)} \left[{4d \over a_4^{(C)}}\right]^{1/3}
\end{equation}
associated with this $c_5=4 d c_4^{(C)}$ coefficient makes 
$a_5^{(C)}=717  \nm.$ Figure
\ref{fig:example_sticking_probabilities_loglog}
shows a slight decrease in the
sticking probabilities, the effect being evidently less here than 
in the case of the thick slab.

\subsection{Hydrogen on `thick' Helium}
Rather atypical, but extremely favourable parameters 
($c_3=18 \; \mev \; \ang^3$) are found
in the case of Hydrogen atoms incident on bulk liquid Helium.
Evidence for quantum reflection
was experimentally seen in this system. \cite{doyle}
A comparison 
with the parameters used in our example of Ne on SiN:
\begin{equation} 
m_{\rm Ne}/m_H=20.03 \hspace{.2cm}
\mbox{and} \hspace{.2cm}
c_3^{({\rm Ne-SiN})}/c_3^{({\rm
    H-He})}=220/18. 
\end{equation}
With the use of Eq. (\ref{T_equation}), we see
that the sticking probabilities for
this case are in fact the same curves as in
Figure
\ref{fig:example_sticking_probabilities_loglog}
except shifted to the right in temperature by about 6.1 orders of
magnitude.  This puts it exactly in the milli-Kelvin regime where 
sticking probabilities of about 0.01 to 0.03 were observed
as temperatures ranged from about 0.3 mK to 5 mK. \cite{doyle}
However, the sticking probabilities predicted by the 
`semi-infinite slab$^{(C)}$' curve of Fig.
\ref{fig:example_sticking_probabilities_loglog}
are about a factor 2.5 too large, but we feel there is good reason
for this.  We already mentioned the qualitative manner in which the 
Casimir forces were included but it seems that a greater error is
caused for another reason.  The length scale $a_3 = 17$ nm for H-He
is so small that the WKB error is close in (see Fig. \ref{fig:wkb_error}) where
the interaction potential is not exactly of the form $\sim 1/x^3$.
Practically speaking this means that the region over which we 
must integrate Eq.(\ref{1d_equation}) must include points 
close to the slab to get some convergence and thus we are 
violating the assumption that the potential is $\sim 1/x^3$ there.
This problem would not plague the Ne-SiN case too much, because the
length scale there is substantially bigger.  For H-He we must include
some short-range information to get an improvement.  Still it is the
long-range forces that are mostly responsible.  
Ref \cite{cole1,amsterdam} and others have modeled this 
close range behaviour and obtained better agreement; the improvement
coming from explicit consideration of the bound states supported
by the close range potential. 
These appear in the  potential matrix elements of perturbation theory.
 
\section{Relation to threshold behavior}
\label{relation_to_threshold_behaviour}
We now wish to take a broader view of the quantum reflection behaviour
at threshold ($k \rightarrow 0 $), and the sticking that sets in as
the energy is increased - a Post Threshold behavior.  In particular we
want to make connection to, and extend the well-known threshold
behaviors of inelastic rates which were first stated most generally by
Wigner in Ref \cite{Wigner}.  For example,  Wigner showed that
the exothermic excitation rates for collisions between two bodies with
bound internal degrees of freedom tend to a constant value as their
relative translational energy tends to 0, provided there is no
resonance at the 0 translational energy threshold.
Equivalently, the exothermic inelastic cross-section diverges as $1/v $, a
fact known in the still
older literature as the `$1/v $ law'. $ v $
is the relative velocity of the collision.
Notice especially the proviso in the statement above, that there be no
resonance at the threshold energy; suggesting that the many
resonances between 0 and $\epsilon$ provided by a many body target
could make the law inoperative.
But the entire thrust of Part 1 was to establish quite generally that
this many-resonance
regime was precisely the one for which the old `$1/v $ law'
is reinstated.

Here we re-examine the Wigner behaviour from a
different point of view using our understanding of quantum
reflection. 
In addition to furthering an intuitive understanding of
the Wigner behaviour,  viewing things in this way will lead naturally to
predicting a generic post threshold behavior (e.g. the 1/v law is replaced
by a $ 1/v^2$ law) and an understanding of when the sticking sets back in
as $\epsilon $ is increased.

The reader will have noted that we have shifted our attention to a
three dimensional geometry of incidence on a localized cluster instead
of
the one dimensional case of incidence on a slab.  
So long as the target dimensions are dwarfed by the incidence
wavelength we will find that both problems are effectively one
dimensional
due to the fact that it is only the s-wave which can penetrate the
interaction region.
For clarity we will deal with both cases separately.

\subsection{Threshold and Post-Threshold Inelastic Cross-sections}
The starting point is the template provided by the golden rule 
\begin{equation}
 d \sigma_{e \rightarrow c} \propto {1 \over k} \rho(E_c) \left|
\int\limits_{\rm all\ \vec r} d^3 r \ \phi^{(-)}_{c,\vec k_c}(\vec r) \
  U_{ce}(\vec r)  \phi^{(+)}_{e,\vec k}(\vec r)
\right|^2
\label{golden_rule_sigma}
\end{equation}
for the differential cross-section for inelastic transitions from
internal state $\Omega_e(u)$ to $\Omega_c(u)$
 where $\vec k$ and $\vec k_c$
 are the incoming and
outgoing directions of the incident atom.  We describe briefly how
Eq. (\ref{golden_rule_sigma}) is arrived at.

For each internal state $\Omega_c(u)$ $(c=1,2,\cdots n$) that we may
imagine freezing the target in ($u$ incorporates all the target
degrees of freedom),
there is some effective potential felt by the incoming atom.  These
potentials are just the diagonal elements of the complete interaction
potential $U(x,u)$ in the $\Omega_c(u)$ basis, which if present all by
themselves (off-diagonal elements 0) could only cause an elastic collision
to occur. It is the off-diagonal elements that
may be thought of as causing the inelastic transitions. Treating them as a
perturbation on the elastic scattering
wavefunctions we use the Golden Rule to obtain
Eq.(\ref{golden_rule_sigma}).
$\rho(E)$ is the energy density of states of the free atom.
$\phi^{(+)}_{e,\vec k}(\vec r)$ is the entrance channel wavefunction
and
$\phi^{(-)}_{c,\vec k_c}(\vec r)$ is the 
 final
channel wavefunction.
They are both exact elastic scattering wavefunctions in the  
 potentials  $U_{ee}(\vec r)$ and
 $U_{cc}(\vec r)$ respectively.
The factor of $1/k$  divides the Golden Rule rate by
the flux to get the probability. 

Now all the $k$ dependence of $d \sigma_{e \rightarrow c}$ and hence 
$\sigma_{e \rightarrow c}$ is due to \newline
1) the factor $1/k$ and \newline
2) The sensitive $k$-dependence of the amplitude of the entrance
channel wavefunction inside the interaction region over which the
overlap integral of Eq.(\ref{golden_rule_sigma}) takes place.  This is
simply because the incoming amplitude is more reflected
away 
by the potential as $k \rightarrow 0$ resulting in the interior
amplitude being suppressed by a factor of $k$ as compared to 
what one would expect classically.

\subsubsection{Incidence on a Slab}
\label{incidence_on_slab}
For this one-dimensional situation we speak of an inelastic
probability instead of a cross-section, but otherwise 
Eq.(\ref{golden_rule_sigma}) remains entirely valid here also with the
obvious modifications.

For $k\rightarrow 0$, when WKB is invalid, we established quite
generally \cite{proposition_paper1} that the entrance channel wavefunction $\phi_e(x) $ when
normalized to have a fixed incoming flux, had its amplitude inside the
interaction region behaving like
\begin {equation}
\phi_e(x_{\rm inside}) \sim k        \hspace{2cm}   {\rm Threshold}
\label{threshold_phi_1d}
\end{equation}
Now the change from quantum reflection at threshold to sticking at
post threshold (see Fig.2) begins to a occur at  those energies at which the
WKB wavefunctions - which show no quantum reflection - may be
increasingly trusted.
At these energies where WKB is valid we may simply use the well-known
WKB amplitude factor $1/\sqrt{k(x)}$, to conclude that
\begin{equation}
\phi_e(x_{\rm inside}) \sim \sqrt{k} \hspace{2cm}   {\rm Post-Threshold}.
\label{post_threshold_phi_1d}
\end{equation}
The probability density of being found inside then behaves like $k^2$ at
threshold (quantum reflection) and like $k$ at post threshold (no quantum
reflection) respectively.

It is quite natural that the probability density inside the interaction
region is smaller 
compared to
the outside by a factor of $k$, even when there is no quantum
reflection. This is simply a kinematical effect:
 where the particle is moving
faster, it is less  likely to be
found by a factor inversely proportional to its velocity there.
Classically what is unexpected is that for
 small enough $k$ near threshold, the probabilities  inside are {\em
further}  suppressed by a
factor of $k$.  Quantum reflection of the amplitude from the region around
$|V(x)|=\epsilon$ (section \ref{quantum_reflection_and_WKB}), goes hand in
hand with the quantum suppression of the amplitude within this region.
 So finally including this $k$-dependence of the
amplitude of $\phi_e(x)$ found in equations (\ref{threshold_phi_1d})
 and (\ref{post_threshold_phi_1d})
we get
\begin{eqnarray}
&P&_{e \rightarrow c} \propto k      \hspace{2cm} {\rm Threshold} \nonumber
\\
&P&_{e \rightarrow c} \propto const. \hspace{1.0cm} {\rm Post-Threshold}
\label{1_d_probabilities}
\end{eqnarray}

\subsubsection{Incidence on a cluster}
Since for large
wavelengths 
only the s-wave interacts
with the cluster it is clear that the problem may be reduced in the
usual manner to a one dimensional problem again. Therefore for a unit
{\em s-wave flux}
the inelastic probabilities will behave as before as in equations
(\ref{1_d_probabilities}), but  what is really relevant is a unit
{\em plain wave } flux which
provides a s-wave flux of $\pi / k^2$.  i.e. Even though the problem 
is one-dimensional
in the radial co-ordinate, the required normalization for the incoming 
flux is not fixed to be a constant as before, but is now required to grow
as $\sim 1/ k^2$, in order to correctly account for the increasing 
(as $k \rightarrow 0$)
range of impact parameters that all `count as' s-wave.
Thus we have simply to multiply the one-dimensional 
probabilities of equations (\ref{1_d_probabilities}) by this factor
of $ 1/k^2$, and conclude that the 
 inelastic cross-sections for this cluster geometry behave like
\begin{eqnarray}
&\sigma&_{e \rightarrow c} \propto {1 \over k} \hspace{2cm} {\rm
  Threshold}  \nonumber \\
&\sigma&_{e \rightarrow c} \propto {1 \over k^2} \hspace{1.0cm} {\rm
Post-Threshold}
\label{3d_cross_sections}
\end{eqnarray}
The Threshold result of Eq.~(\ref{3d_cross_sections}) is just the
Wigner `1/v law' we spoke of in section
\ref{relation_to_threshold_behaviour}. But now we can say more.
 As the incoming wavelength $\lambda $ increases, we first witness
for large enough $\lambda $ a quadratic dependence to the exothermic
cross-section ($\sigma \propto \lambda^2 $).
 It is only at still larger wavelengths that this dependence eventually
changes over
 to a linear one ($\sigma \propto \lambda$).  This
 happens when the sticking yields to the quantum reflection.  This
 energy is mostly determined
  by  the long range form of the potential, and has nothing to
 do with the bound state energies or any other details involving the
 interaction potential.

\section{Conclusion}

Examining the WKB error term provided a quick and easy way to estimate
the threshold temperatures required to observe quantum reflection. 
 It became transparent that only power laws dying faster than $-1/x^2$
 were 
capable of acting as quantum
reflectors.  The validity of WKB at higher
temperatures heralded a post-threshold behavior in which the atom
sticks.
Even for
other geometries
 such as incidence on a localised three dimensional cluster, a WKB
 analysis together with the Fermi Golden Rule provided a simple
 understanding of this
 threshold and post-threshold behavior in terms of inelastic
processes being shut off due to a reflection of the incoming
amplitude.  The extremely long incoming wavelength is invariably
impedance  mismatched (for potentials shorter
ranged than $1/r^2$) by the abrupt change of wavelength in the
interaction region,
 and is therefore reflected.
It should be clear that even a repulsive interaction will obviously provide
such a
mismatch so that the Wigner behavior, or quantum reflection, is quite
general; though of
course most dramatic if the potential is attractive as we have been
considering throughout.

This effect of quantum reflection/suppression, which is ultimately
responsible for the threshold behaviour, is   dynamical   in
that it is caused by the presence of the interaction potential.
We feel that the original derivation by Wigner
that focuses on the $k\rightarrow 0$ behaviour of the free space wave
functions
tends to obscure this  physical origin of threshold behaviour.
The golden rule approach makes it more explicit and especially paves
the way for predicting the Post-Threshold behaviour.

\acknowledgements

A.M. is most grateful to Michael Haggerty for his kind help and advice
and for 
always being
available to discus things with. A.M. thanks  Alex Barnett for
pointing him to the 
references on the Casimir interactions.

This work was supported by the National Science Foundation
through a grant for the Institute for Theoretical Atomic and
Molecular Physics at Harvard University and Smithsonian
Astrophysical Observatory:National Science Foundation Award Number
CHE-0073544.

This work was also supported by the National Science Foundation by 
grants PHY-0071311 and PHY-9876927.


\begin{references}
\bibitem{proposition_paper1} We gave the heuristic elementary reason for 
the suppressed behaviour 
of Eq.~(\ref{threshold_phi_1d}) 
in Sec III of paper I,where we then attempted to rigorously justify it.
\end{references}
\end{document}